\documentstyle[12pt,epsfig]{article}
\begin{document}
\newcommand{\newc}{\newcommand}
\newc{\R}{$R$}
\newc{\charginom}{M_{\tilde \chi}^{+}}
\newc{\mue}{\mu_{\tilde{e}_{iL}}}
\newc{\mud}{\mu_{\tilde{d}_{jL}}}
\newc{\beq}{\begin{equation}}
\newc{\eeq}{\end{equation}}
\newc{\barr}{\begin{eqnarray}}
\newc{\earr}{\end{eqnarray}}
\newc{\ra}{\rightarrow}
\newc{\lam}{\lambda}
\newc{\eps}{\epsilon}
\newc{\gev}{\,GeV}
\newc{\eq}[1]{(\ref{eq:#1})}
\newc{\eqs}[2]{(\ref{eq:#1},\ref{eq:#2})}
\newc{\etal}{{\it et al.}\ }
\newc{\Hbar}{{\bar H}}
\newc{\Ubar}{{\bar U}}
\newc{\Dbar}{{\bar D}}
\newc{\Ebar}{{\bar E}}
\newc{\eg}{{\it e.g.}\ }
\newc{\ie}{{\it i.e.}\ }
\newc{\nonum}{\nonumber}
\newc{\lab}[1]{\label{eq:#1}}
\newc{\lle}[3]{L_{#1}L_{#2}\Ebar_{#3}}
\newc{\lqd}[3]{L_{#1}Q_{#2}\Dbar_{#3}}
\newc{\udd}[3]{\Ubar_{#1}\Dbar_{#2}\Dbar_{#3}}
\newc{\dpr}[2]{({#1}\cdot{#2})}
\newc{\rpv}{{\not \!\! R_p}}
\newc{\rpvm}{{\not \! R_p}}
\newc{\rp}{$R_p$}
\newc{\gsim}{\stackrel{>}{\sim}}
\newc{\lsim}{\stackrel{<}{\sim}}

\newcommand{\chargino}{{\tilde \chi}^{+}}
\newcommand{\charginol}{{\chi}^{+}_{l}}
\newcommand{\neutralino}{{\tilde \chi}^{0}}

\newcommand{\ffig}[4]{\begin{figure}[htbp]\vfill\begin{center}
\mbox{\epsfig{figure=#1,height=#2}}\caption{#3}\label{#4}
\end{center}\vfill\end{figure}}
\newcommand{\wfig}[4]{\begin{figure}[htbp]\vfill\begin{center}
\mbox{\epsfig{figure=#1,width=#2}}\caption{#3}\label{#4}
\end{center}\vfill\end{figure}}
\setcounter{figure}{0}

\title{Chargino Pair Production at LEP2 with Broken R-Parity: 4-jet
Final States}
\author{H. Dreiner$^1$, S. Lola$^2$ and P. Morawitz$^3$}
\date{{\small $^1$ Rutherford Lab., Chilton, Didcot, OX11 0QX, UK\\
$^2$ CERN-TH, CH-1211 Geneve 23\\
$^3$ Imperial College, HEP Group, London SW7 2BZ, UK}}
\maketitle


\begin{abstract}
\noindent We study the pair production of charginos in $e^+e^-$-collisions 
followed by the decay via R-parity violating $LQ\Dbar$ operators. We determine
the complete matrix element squared for chargino decays via $LQ\Dbar$ or
$LL\Ebar$ operators. We find regions in MSSM  parameter space where
the chargino mass is $52.5\gev$ and the R-parity violating decays of the 
charginos dominate the gauge decays to neutralinos. At LEP2 this then 
leads to additional 4 jet events which could explain the excess recently 
observed by ALEPH. 

\vspace{5cm}
\end{abstract}

\begin{center}
{\it Submitted to Physics Letters B.}
\end{center}
\newpage

\section{Introduction}
Supersymmetry predicts many new particles with thresholds possibly
within reach of the LEP2 collider at CERN \cite{greg}.  Chargino pair
production is a promising candidate for a first signal. It has been widely
studied \cite{king4} within the minimal supersymmetric standard model (MSSM),
where R-parity $R_p=(-)^{2S+3B+L}$ is conserved\footnote{Here S: spin, B:
baryon number, L: lepton number.}. There the chargino cascade decays to the
lightest neutralino which is stable and escapes detection. The signal is partly
characterised by missing transverse momentum. 

From a theoretical point of view it is just as likely that $R_p$ is violated
($\rpv$) \cite{rpv-theory,symms}. The lightest supersymmetric particle is then
no longer stable. If it decays within the detector the missing transverse
momentum signal is diluted and the signal has other main characteristics
\cite{rppheno}. Direct searches for lepton signals in ALEPH at LEPI energies
have previously addressed this issue \cite{ALEPHLEPIrpv}, and have shown that the SUSY limits
obtained from searches under the assumption of $R_p$ conservation also hold
under the simultaneous violation of R-parity and lepton number.
It is the purpose of this letter to study the production and
decay of charginos at LEP2 with broken R-parity with particular emphasis on
4-jet final states. 

Recently ALEPH observed anomalously large 4-jet production at LEP2 while
running at $\sqrt{s}=133-136\gev$ \cite{4jets}. There have been several
proposed solutions \cite{solutions} in particular two \cite{rpsol1,rpsol2}
which also propose mechanisms within supersymmetry with broken R-parity. Of the
latter two, the first considers pair production of scalar sneutrinos and their
subsequent decay via an $LQ\Dbar$ operator. The second considers squark pair
production followed by the decay via a $\Ubar\Dbar\Dbar$ operator. We present
here a third possible explanation via broken R-parity namely the production and
decay of charginos. As we see below, this is experimentally distinguishable and
relies on the  $LQ\Dbar$ operator. 

When R-parity is broken the superpotential contains the additional baryon- 
and lepton-number violating Yukawa couplings\footnote{Here 
$L$: lepton $SU(2)$ doublet superfield, $Q$: quark doublet superfield,
$\Ebar$: charged lepton singlet superfield, $\Dbar$:
down-like quark singlet superfield, $\Ubar$: up-like quark singlet superfield. 
$i,j,k$ are generation indices. $\lam,\lam',\lam''$ are dimensionless Yukawa 
couplings.} \cite{weinberg} 
\beq
W_{\rpvm} =\lam_{ijk}L_iL_j\Ebar_k + \lam'_{ijk} L_iQ_j\Dbar_k
+\lam''_{ijk}\Ubar_i\Dbar_j \Dbar_k.
\label{eq:rpar}
\eeq
The superpotential contains 45 operators; combinations of the lepton- and
baryon-number violating couplings can lead to proton decay in disagreement with
the experimental bounds \cite{sher}. Thus some symmetry must be imposed which
prohibits a subset of the terms. Several examples have been considered in the
literature \cite{symms}. In most models motivated by unification (including
gravity), there is a preference for allowing the lepton number violating terms
over the baryon number violating terms. In addition, the strictest laboratory
bounds are on the lowest generation $\Ubar_i\Dbar_j\Dbar_k$ operators, {\it
e.g.} $\lam''_{121}<10^{-6}$ \cite{goity} rendering them unimportant for
collider searches. It is difficult to construct models which allow for large
higher generation couplings $\lam''$ and which still satisfy this strict bound
on $\lam''_{121}$ since the quark mixing is known to be non-zero. In our
specific example below, we shall thus focus on the case of a single dominant
$L_iQ_j\Dbar_k$ operator. The present experimental bounds on these operators
are given in Table \ref{lambdabounds}. The $LQ\Dbar$ operators do not affect
chargino production but can significantly alter the decay patterns of the
charginos. As we show below, in relevant regions of parameter space the
R-parity violating decay of the chargino dominates. These decays then lead to
four jet final states which could explain the discrepancy observed by ALEPH.

\begin{table}
\begin{center}
\begin{tabular}{|cc|cc|cc|}\hline
$\lam'_{111}$ &0.0004& $\lam'_{211}$ &0.09&$\lam'_{311}$& 0.14\\ \hline
$\lam'_{112}$ &0.03& $\lam'_{212}$ &0.09&$\lam'_{312}$& 0.14\\ \hline
$\lam'_{113}$ &0.03& $\lam'_{213}$ &0.09&$\lam'_{313}$& 0.14\\ \hline
$\lam'_{121}$ &0.26& $\lam'_{221}$ &0.18&$\lam'_{321}$& -\\ \hline
$\lam'_{122}$ &0.45& $\lam'_{222}$ &0.18&$\lam'_{322}$& -\\ \hline
$\lam'_{123}$ &0.26& $\lam'_{223}$ &0.18&$\lam'_{323}$& -\\ \hline
$\lam'_{131}$ &0.26& $\lam'_{231}$ &0.44&$\lam'_{331}$& 0.26\\ \hline
$\lam'_{132}$ &0.51& $\lam'_{232}$ &0.44&$\lam'_{332}$& 0.26\\ \hline
$\lam'_{133}$ &0.001& $\lam'_{233}$ &0.44&$\lam'_{333}$& 0.26\\ \hline
\end{tabular}
\end{center}
\caption{{\footnotesize  Bounds on the Yukawa couplings of the $L_iQ_j\Dbar_k$
operators {\protect \cite{gautam}}. The bounds are all to be multiplied by a
scalar fermion mass ${\tilde M}_0 /100 \gev$ except for the bound on
$\lam'_{111}$ which also depends on the gluino mass {\protect \cite{hirsch}}. 
There are no bounds on the  couplings $\lam'_{32k}$. \label{lambdabounds}}}
\end{table}

\section{Chargino Decays}
\subsection{SUSY Spectrum}
The decay pattern of the chargino depends foremost on the supersymmetric
spectrum. In low energy supersymmetry\footnote{For a review see
\cite{haber,nilles}.}, when $SU(2)_L\times U(1)_Y$ has been broken to
$U(1)_{em}$ the gauginos mix with the Higgsinos to form the chargino and
neutralino mass eigenstates. The masses depend on the $SU(2)_L$, $U(1)_Y$
gaugino masses $M_2$, and $M_1$, the Higgs mixing parameter $\mu$ and the ratio
of the vacuum expectation values $\tan\beta$. For any fixed values of these
parameters, we can determine the gaugino spectrum completely. In particular, we
can determine the nature of the lightest supersymmetric particle (LSP), and
which gaugino decay modes are kinematically accessible to the chargino. 

In grand unified theories $M_2$ and $M_1$ are related by $M_1=\frac{5}{3}\tan^2
\theta_W M_2$. We shall impose this constraint throughout this letter and we
thus only have 3 free parameters in the gaugino sector: $(M_2,\,\mu,\,\tan\beta
)$. In Fig. \ref{model} we have fixed $\tan\beta=2,35$. For $\tan\beta=2$ the
black band denotes the range of parameters where the chargino is the LSP. In
this range the chargino mass is never above $\approx23\gev$. For $M_{\chi^+}
<(M_{Z^0}/2)$ the chargino can contribute to the $Z^0$\  width. This is
independent of the chargino decay and therefore does not depend on whether
R-parity is conserved or not. LEP1 measurements of the $Z^0$width have
determined a model independent lower bound on the mass of the chargino of 
$M_{\chi^+}\geq45.2\gev$ \cite{lep}. This is included in Fig.\ref{model} as a
narrow dashed curve.  We see that within the SUSY-GUT framework, given the
experimental constraints, the chargino can not be the LSP. For $\tan\beta=35$
there is also a band where the chargino is the LSP but it is very small and
below the resolution of Fig.\ref{model}. It is excluded by the LEP1 measurement
as well.

\subsection{Matrix Element}
We now study the $\rpv$ chargino decays for the explicit case of a $L_iQ_j
\Dbar_k$ operator. As we discuss in the appendix, the results can be easily
translated to the decay via a $L_iL_j\Ebar_k$ operator. A positively charged
chargino\footnote{The index l=1,2 denotes the two chargino mass eigenstates,
respectively. $l=1$ is the lighter of the two. Indices $i,j,k$ are generation
indices.} ${{\tilde\chi}^+}_l$  can decay to the following final states 
\barr 
{{\tilde\chi}^+}_l \ra\left\{ \begin{array}{l}
{\nu}_i+u_j+{\bar d}_{kR}   \,\,\,\,\,\,\,\,   (\ref{eq:finalstates}.1)   \\ 
e_i^++{\bar d}_j+{ d}_{kR}      \,\,\,\,\,\,  (\ref{eq:finalstates}.2) \\ 
e_i^++{\bar u}_j+{ u}_{kR}      \,\,\,\,\,\,  (\ref{eq:finalstates}.3) \\ 
{\bar \nu}_i+{\bar d}_j+{u}_{kR} \,\,\,\,\,\,\,\, (\ref{eq:finalstates}.4)
\end{array}
\right.
\label{eq:finalstates}
\earr
\wfig{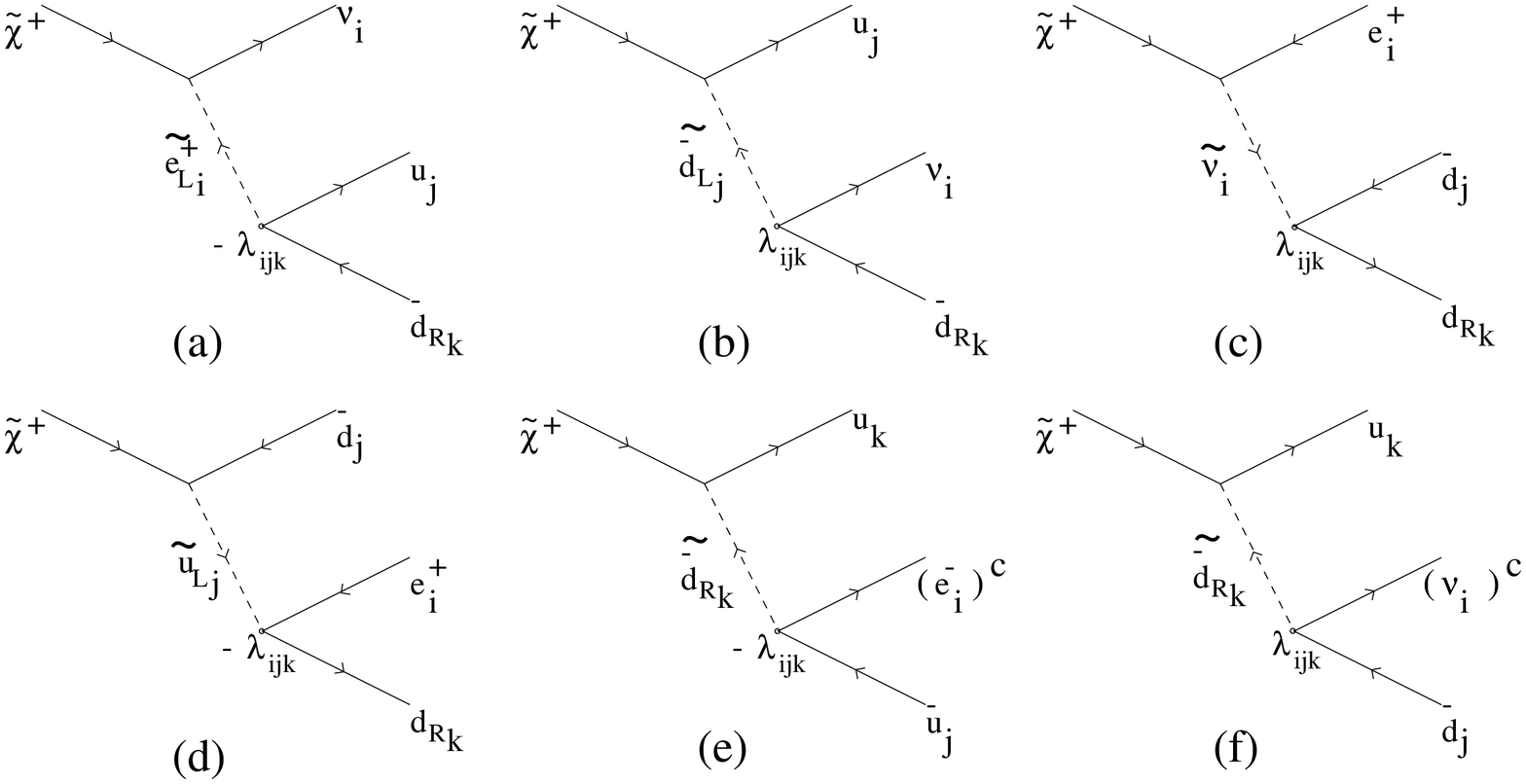}{12cm}{\footnotesize
  Feynman diagrams for direct $\rpv$ Chargino decays via the $L_iQ_j{\bar D}_k$ operator.}{diags}
Figs.(\ref{diags}a)-(\ref{diags}f) show the Feynman diagrams for chargino
decays into the final states (\ref{eq:finalstates}.1)-(\ref{eq:finalstates}.4).
The $\rpv$ vertices are labelled by the Yukawa coupling strength ($\lam'_{ijk}
$)\footnote{Some of the vertices are labelled by $-\lam'_{ijk}$, this is
because the proper $SU(2)$ invariant is $\eps_{ab} L_i^aQ_j^b\Dbar_k$, where $
\eps_{01}=-\eps_{10}$ and $\eps_{00}=\eps_{11}=0$. In Eq.(\ref{eq:rpar}) we 
have suppressed the $SU(2)$ indices $a,b$.}. Note that there are two diagrams
(\ref{diags}a,b) and (\ref{diags}c,d) for the final states
(\ref{eq:finalstates}.1) and (\ref{eq:finalstates}.2) respectively. 
For the four decay modes the amplitudes squared are given by
\barr
\label{eq:matrix1}
|{\cal M}_1|^2 &=& 4n_cg^2{\lam'}^2 \left[
\frac{\alpha_R^2}{R^2({\tilde e}_{iL})} \dpr{\charginol}{\nu_i} 
                     \dpr{u_j}{{\bar d}_k} \nonumber\right. \\
&&+ \frac{\dpr{\nu_i}{{\bar d}_k}}{R^2({\tilde d}_{jL})}
\left\{(\beta_L^2+\beta_R^2)\dpr{{\charginol}}{u_j}
+2{\cal R}e(\beta_L\beta_R^*m_{uj}M_{{\charginol}})\right\}\nonumber \\
&&\left.-{\cal R}e\left\{
 \frac{\alpha_R}{R({\tilde e}_{iL})R({\tilde d}_{jL})}\left(\beta_L^*
m_{uj} M_{{\charginol}}
\dpr{u_j}{{\bar d}_k}+\beta_R^*{\cal G}(p,\nu_i,{\bar d}_k,u_j)\right)
\right\} \right]
\\
\label{eq:matrix2}
|{\cal M}_2|^2 &=&  
4n_cg^2{\lam'}^2 \left[\frac{\dpr{d_j}{{\bar d}_k}}{R^2({\tilde \nu}_{iL})}
\left\{(\gamma_L^2+\gamma_R^2)\dpr{{\charginol}}{e_i} 
+2{\cal R}e(\gamma_L\gamma_R^* m_{ei}M_{{\charginol}}
)\right\} \nonumber \right.\\
&& + \frac{\dpr{e_i}{{\bar d}_k}}{R^2({\tilde u}_{jL})}
\left\{(\delta_L^2+\delta_R^2)\dpr{{\charginol}}{d_j} 
+2{\cal R}e(\delta_L\delta_R^* m_{dj}M_{{\charginol}})
\right\}\nonumber \\
&&\left.-\frac{1}{R({\tilde \nu}_{iL})R({\tilde u}_{jL})} {\cal R}e\left\{
\gamma_L \delta_L^*{\cal G}({\charginol},e_i,{\bar d}_k,d_j)
+\gamma_L \delta_R^* m_{dj}M_{{\charginol}}\dpr{e_i}{{\bar d}_k} \right.\right.
\nonumber \\
&& \left.\left.
+\gamma_R \delta_L^* m_{ei}M_{{\charginol}}\dpr{d_j}{{\bar d}_k}
+\gamma_R \delta_R^* m_{ei}m_{dj}\dpr{{\charginol}}{{\bar d}_k} \right\}\right]
\\
\label{eq:matrix3}
|{\cal M}_3|^2 &=& 
\frac{2n_c{\lam'}^2g^2m_{dk}^2 |U_{l2}|^2}{M_W^2\cos^2\beta R^2({\tilde
d}_{kR})}\dpr{e_i}{u_j}
\dpr{{\charginol}}{u_k}\\
\label{eq:matrix4}
|{\cal M}_4|^2
&=& \frac{2n_c{\lam'}^2g^2m_{dk}^2|U_{l2}|^2}{M_W^2\cos^2\beta R^2({\tilde
d}_{kR})}\dpr{\nu_i}{d_j} \dpr{{\charginol}}{u_k}
\earr
$\alpha_{L,R},\beta_{L,R},\gamma_{L,R}$ and $\delta_{L,R}$ are couplings and
are given in the appendix. The final-state momenta are denoted by the particle
symbols. $M_{{\charginol}}$ is the chargino mass and $m_{ei,dj,dk}$ are the
final state fermion masses. $n_c=3$ is the colour factor. The function ${\cal
G}(a,b,c,d)= \dpr{a}{b}\dpr{c}{d}-\dpr{a}{c}\dpr{b}{d}+ \dpr{a}{d} \dpr{b}{c}
$. The square of the propagators $R(p)$ are given in the appendix. We have
included all mass effects. In most applications $\beta_L,\gamma_R,\delta_R
\approx0$. For $j=3$, $\beta_L$ is not negligible but the decay
$\chi^+\ra\nu_i+t+{\bar d}_{kR}$ is kinematically prohibited unless $M_
{\chi^+}>m_{top}$. For large $\tan\beta$ and $i,j=3$ $\gamma_R,\delta_R$ can be
important. Note that the last two decay modes are proportional to
$(m_{dkR}/(\cos\beta M_W))^2$ and are thus suppressed in most of the parameter
space. The analogous decay of the neutralino has been given in
\cite{peterherbi}. The partial widths are given by the integration over phase
space 
\beq
\Gamma_l= \int \frac{1}{2\pi^3}\frac{1}{16 M_{\charginol}}|{\cal M}_l|^2 
dE_idE_j
\eeq
and we have averaged over the initial spin states of the chargino. If we
neglect the final state masses the integrals can be performed analytically.
We find
\beq
\Gamma_1({{\tilde \chi}^+}_l\ra {\nu}_i+u_j+{\bar d}_k) =
4n_cg^2{\lam'}^2 \left[
\alpha_R^2 A(\mue) +
(\beta_L^2+\beta_R^2) A(\mud)+
{\cal R}e
(\alpha_R\beta_R^*) B(\mue,\mud) \right]
\eeq
In the above equation,
\beq
A(\mue) = \frac{1}{32}\left[-5 + 6\mue^2 + (2 - 8\mue^2 +  6\mue^4)
\ln \left ( \frac{\mue^2-1}{\mue^2} \right ) \right]
\eeq
where 
\beq
\mue = \frac{m_{\tilde{e}_i}}{\charginom}
\eeq
are the dimensionless normalized masses. $B(\mue,\mud)$ arises from the
interference term and can be expressed as 
\beq
B(\mue,\mud) = \frac{1}{4}I(\mue,\mud)  + \frac{1}{32} C(\mue,\mud) 
\eeq
where
\beq
I = \int_{0}^{1/2}(x - 2x^2 + x\mud^2 - \mud^4)
        \ln (\mud^2-2x) / (-1 + 2x + \mue^2) dx
\eeq
and
\barr
C(\mue,\mud) &=& -4\mud^2 + (1 - 2\mud^2 - 2\mue^2)\ln(\mud^2)
+ 4(\mud^2 - \mud^2\mue^2)\psi \nonumber \\ 
&&+ 
2(\mud^2 -
     2\mud^4 +   \mue^2 -
     \mud^2\mue^2 -
     \mue^4)\ln(\mud^2)\psi
\earr
Note that in the above, all terms in $A,B$ are normalised with respect to the
chargino mass $\charginom$. $\psi=\ln\{(\mue^2-1)/\mue^2\}$.
For $\Gamma_2$ we find an identical expression, with the only change being in
the masses of the propagators, as one can see from the Feynman diagrams of
Figure 1. When the final state masses are neglected $\Gamma_3=\Gamma_4=0$. 

\subsection{Chargino Width and Lifetime}
If we only consider one non-zero operator $L_iQ_j\Dbar_k$ the $\rpv$
decay width is given by
\beq
\Gamma_\rpvm=\Gamma_1+\Gamma_2+\Gamma_3+\Gamma_4.
\eeq
In the MSSM there are possible further (\rp-conserving) cascade decays of the
chargino ${\tilde\chi}^+_l$ via a virtual W-boson to a lighter neutralino ${
\tilde\chi}^0_m$ 
\barr
\label{eq:mssmcascades1}
\chargino_l &\ra& l^+ + \nu + \neutralino_m, \\
\label{eq:mssmcascades2}
\chargino_l &\ra& q + {\bar q}' + \neutralino_m.
\earr
We denote the MSSM contributions to the chargino width as $\Gamma_{MSSM}$. The
MSSM decay rates have been calculated by various authors \cite{chargino} and
they are functions of the three free parameters ($M_2,\mu,\tan\beta$). The
total chargino width is given by 
\beq
\Gamma_{{\tilde \chi}^+} = \Gamma_\rpvm + \Gamma_{MSSM}.
\eeq
Eventhough the chargino is not the LSP it will nevertheless dominantly decay
directly to an R-parity even final state via the decays
(\ref{eq:finalstates}.1)-(\ref{eq:finalstates}.4) if the ratio $\frac
{\Gamma_\rpvm} {\Gamma_{MSSM}}$ is sufficiently large. This happens in regions
of the MSSM parameter space in which the MSSM cascade decays of the chargino
(\ref{eq:mssmcascades1})-(\ref{eq:mssmcascades2}) are phase space supressed,
i.e. when $M_{{\tilde \chi}^+} \approx M_{{\tilde \chi}^0}$,  {\it and} when
the Yukawa coupling $\lam'_{ijk}$ is not too small. In order to explore the
ratio $\frac{\Gamma_ \rpvm} {\Gamma_{MSSM}}$ numerically we consider a fixed
SUSY {\it scalar} mass spectrum: $m_{{\tilde q}}=500\gev$, ${m_{{\tilde
l}_{L}}} = 200\gev$, ${m_{{ \tilde l}_{R}}} = m_{\tilde \nu} =100\gev$. The
ratio ${m_{{ \tilde l}_{L}}}/m_{\tilde \nu}\geq2$ was chosen to optimize our
4-jet signal. It is consistent with supergravity (SUGRA) models which generally
predict squarks to be the heaviest and sneutrinos and right-handed sleptons to
be the lightest SUSY scalar particles,  but is not a generic feature, \ie there
are regions in {\it SUGRA} parameter space where the ratio is only slightly
greater than 1 \cite{sugra}. A factor of $\sim1.7$ is sufficient for our
argument.

\begin{figure}
\vspace{0.5cm}
\centerline{\hbox{\hspace{2.0cm}
{\psfig{figure=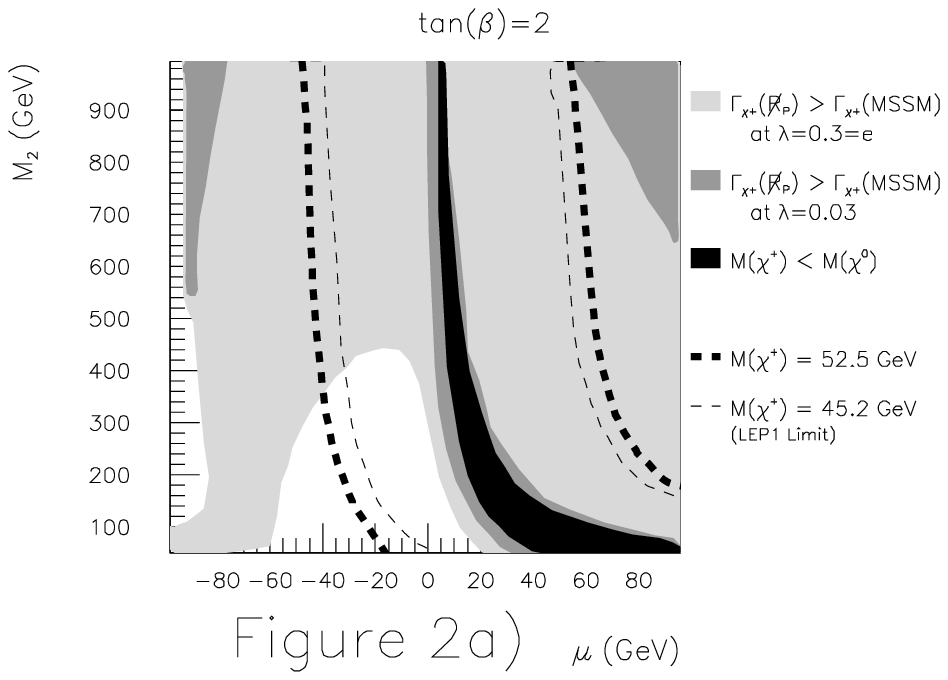,height=10cm,width=10cm}}
{\psfig{figure=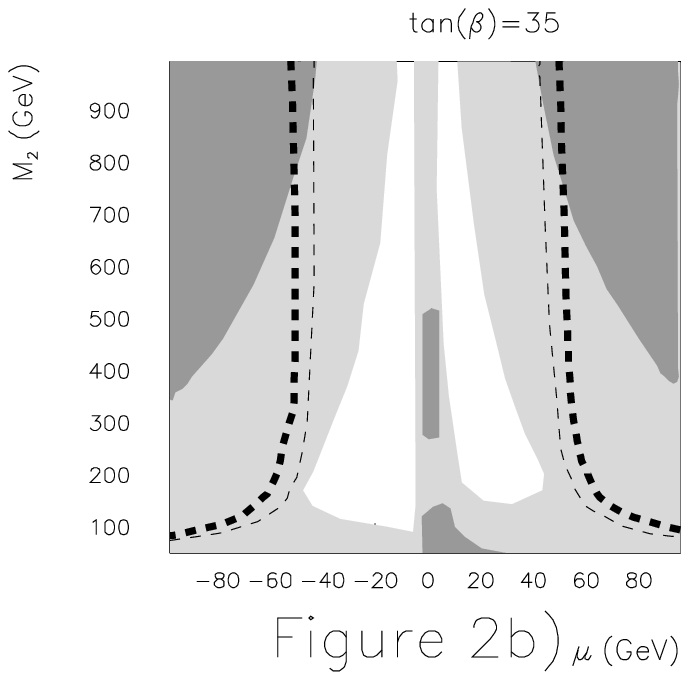,height=10cm,width=10cm}}}}
\vspace{-2.0cm}
\caption{\label{model}\footnotesize  Regions in the $M_2 - \mu$ parameter space in which
$\Gamma_\rpv > \Gamma_{MSSM}$ at $\lam'_{ijk}=0.3$ (light grey area) and
$\lam'_{ijk}=0.03$ (dark grey area), and in which the chargino is the LSP
(black region) for a) $\tan\beta=2$ and b) $\tan\beta=35$. Superimposed are
contours of $M_{\chargino}=52.5\gev$, the region where ALEPH sees an excess in
4-jet events.}
\end{figure}

In Figs.\ref{model}a,b we plot regions in the $M_2 - \mu$ gaugino parameter
space in which $\Gamma_\rpvm > \Gamma_{MSSM}$ for a coupling
strength of $\lam'_{ijk}=0.3$ (light grey area) and $\lam'_{ijk}=0.03$
(dark grey area). As mentioned above, the black area indicates the chargino LSP
region in which the chargino always decays \rp-violating. For the large
Yukawa coupling of $\lam'_{ijk}=e=0.3$ the direct $\rpv$ chargino decays
dominate over the MSSM cascade decays throughout nearly the entire $M_2-\mu$
plane. This is because the MSSM decay to the neutralino is phase space
suppressed whereas the $\rpv$ decay is not coupling suppressed. However, even 
for $\lam=0.03$ there is still a substantial region of parameter space at
large $M_2$ where $\Gamma_\rpvm$ dominates. Here $M_{\chi^+_1}\approx M_
{\chi^0_1}$ {\it and} the $\chi^+\chi^0W^+$ coupling is small. We discuss the phenomenological consequences in more
detail in section \ref{sec:4jets}. 

\wfig{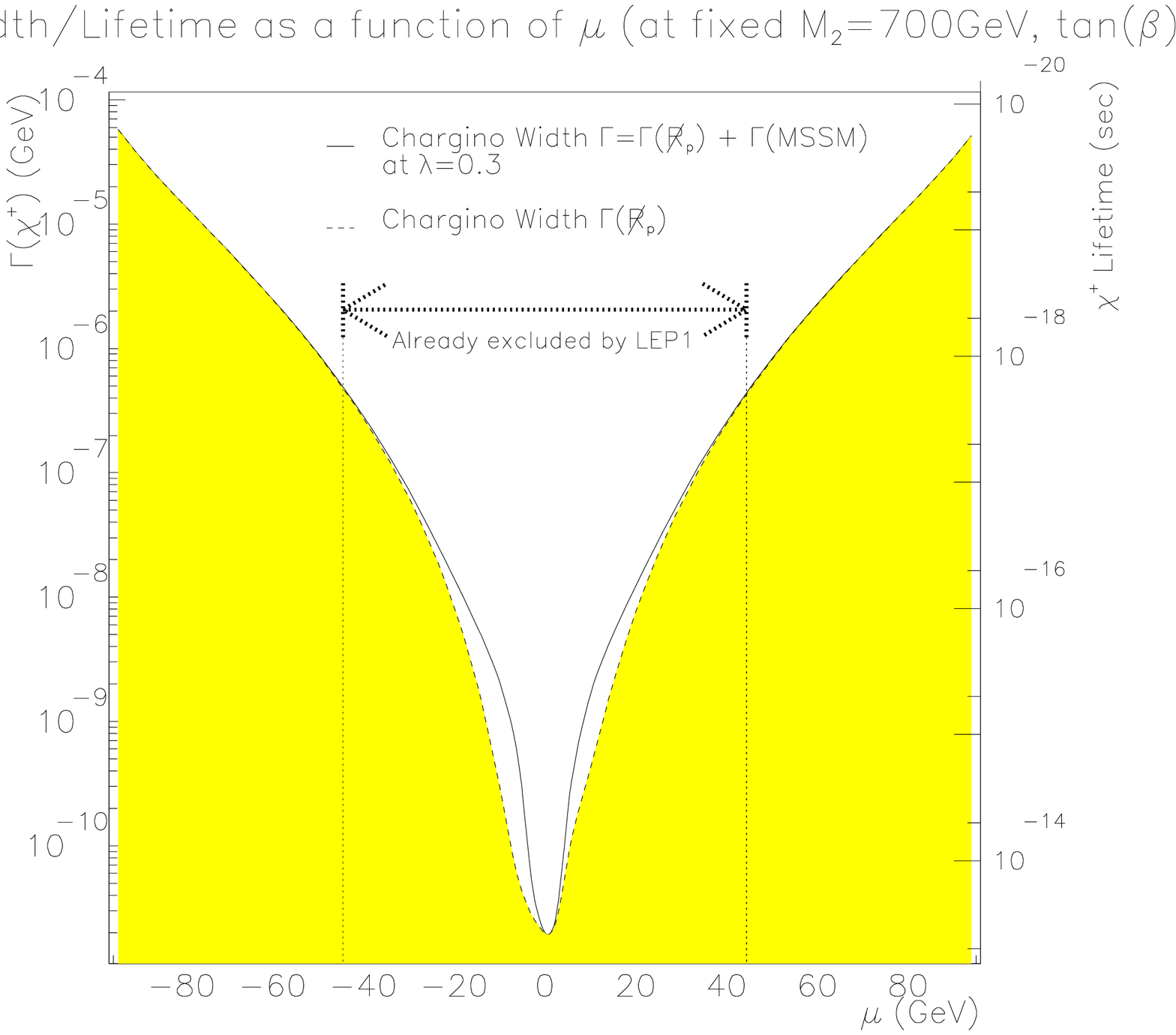}{8cm}{\footnotesize
  Width/Lifetime of the Chargino.}{lifetime}

We now turn to the chargino lifetime. Fig.\ref{lifetime} shows the chargino
width and lifetime as a function of $\mu$ for one particular value of $M_2=
700\gev$ and $\tan\beta=35$. The solid line shows the total width $\Gamma_{
{\tilde \chi}^+} = \Gamma_\rpvm + \Gamma_{MSSM}$ at $\lam'_{ijk}=0.3$, while
the broken line shows the $\rpv$ contribution alone. The latter scales with $
\lam'^2_{ijk}$ as seen in Eqs.(\ref{eq:matrix1}-\ref{eq:matrix4}). So it is
clear that the chargino always decays within the detector. This also holds for
smaller values of $M_2$ and $\tan\beta$.

\subsection{Branching Fractions}
We now determine the branching ratios of the chargino decays into the final
states (\ref{eq:finalstates}.1)-(\ref{eq:finalstates}.4). The decays
(\ref{eq:finalstates}.3) and (\ref{eq:finalstates}.4) are supressed with
respect to (\ref{eq:finalstates}.1) and (\ref{eq:finalstates}.2) by $({\frac
{m_{dk}} {\sqrt 2 M_W \cos\beta }})^2$ because the exchanged virtual
right-handed down-type squark ${\tilde {\bar d_R}}$ (see also diagrams
\ref{diags}e and \ref{diags}f) only couples Higgsino-like to the chargino. The
decays (\ref{eq:finalstates}.1) and (\ref{eq:finalstates}.2) are comparable if
the scalar fermion masses are, {\it i.e.} $m_{{\tilde \nu}}\approx m_{{\tilde
e}}\approx m_{{\tilde u}}\approx m_{{\tilde d}}$. However as pointed out
earlier, the four-jet signal discussed below is enhanced for $m_{{\tilde \nu}}<
m_{{\tilde e}L}$. The ratios of the decay widths are given by 
\beq
{\Gamma}_1 : {\Gamma}_2 : {\Gamma}_3 : {\Gamma}_4 =  ({\frac
{{ m_{\tilde\nu}}} {{m_{{\tilde e}_L}}}}  )^4 : 1 :  
({\frac{7 \times 10^{-6}} {\cos^2 \beta}}) ({\frac {{ m_{{\tilde\nu}}}}
{{{m_{{\tilde d}_R}}} } })^4 :
({\frac{7 \times 10^{-6}} {\cos^2 \beta}}) ({\frac {{ m_{{\tilde\nu}}}}
{{{m_{{\tilde d}_R}}} } })^4.
\eeq 
For our specific model, in which we fix $m_{{\tilde
q}}=500\gev$, $m_{{\tilde l}_L} = 200\gev$, $m_{{\tilde l}_R} = m_{\tilde \nu} =
100\gev$, and $\tan\beta=35$,
\beq
{\Gamma}_1 : {\Gamma}_2 : {\Gamma}_3 : {\Gamma}_4 = 6 \times 10^{-2} : 1 :
10^{-5} :  10^{-5}
\eeq
the decay mode (\ref{eq:finalstates}.2) is dominant over the entire 
$M_2 - \mu$ plane. Thus from now on we neglect the other decay modes
and restrict ourselves to the decay $\chi^+\ra e_i^++{\bar d}_j+d_{Rk}$. This
essential conclusion holds for $m_{{\tilde e}_L}/m_{{\tilde\nu}} \gsim 1.7$.

\subsection {Decay Distributions}
\label{Decay.Distributions}

We consider the energy distributions of the  chargino decay products for the
decay ${\tilde\chi}^+ \ra e_i^++d_j+{\bar d}_k$. The most sizeable effect on
the charged lepton momentum is excerted by the ${\tilde \nu}$-propagator
(Fig.\ref{diags}c). For {$m_{\tilde \nu} \approx M_{\chargino}$} the energy
spectrum of the charged lepton, $E_{e_i}$, is soft. The quark energy spectra $
E_{d_j,{\bar d}_k}$ are much harder. This effect is demonstrated in
Fig.\ref{edis}a. The $E_{e_i}$ spectrum becomes harder the greater the mass
difference $m_{\tilde\nu}-M_{\chi^+}$. However, even for $m_{\tilde\nu}=100\gev
$ it is still substantially softer than the quark spectrum. 

\wfig{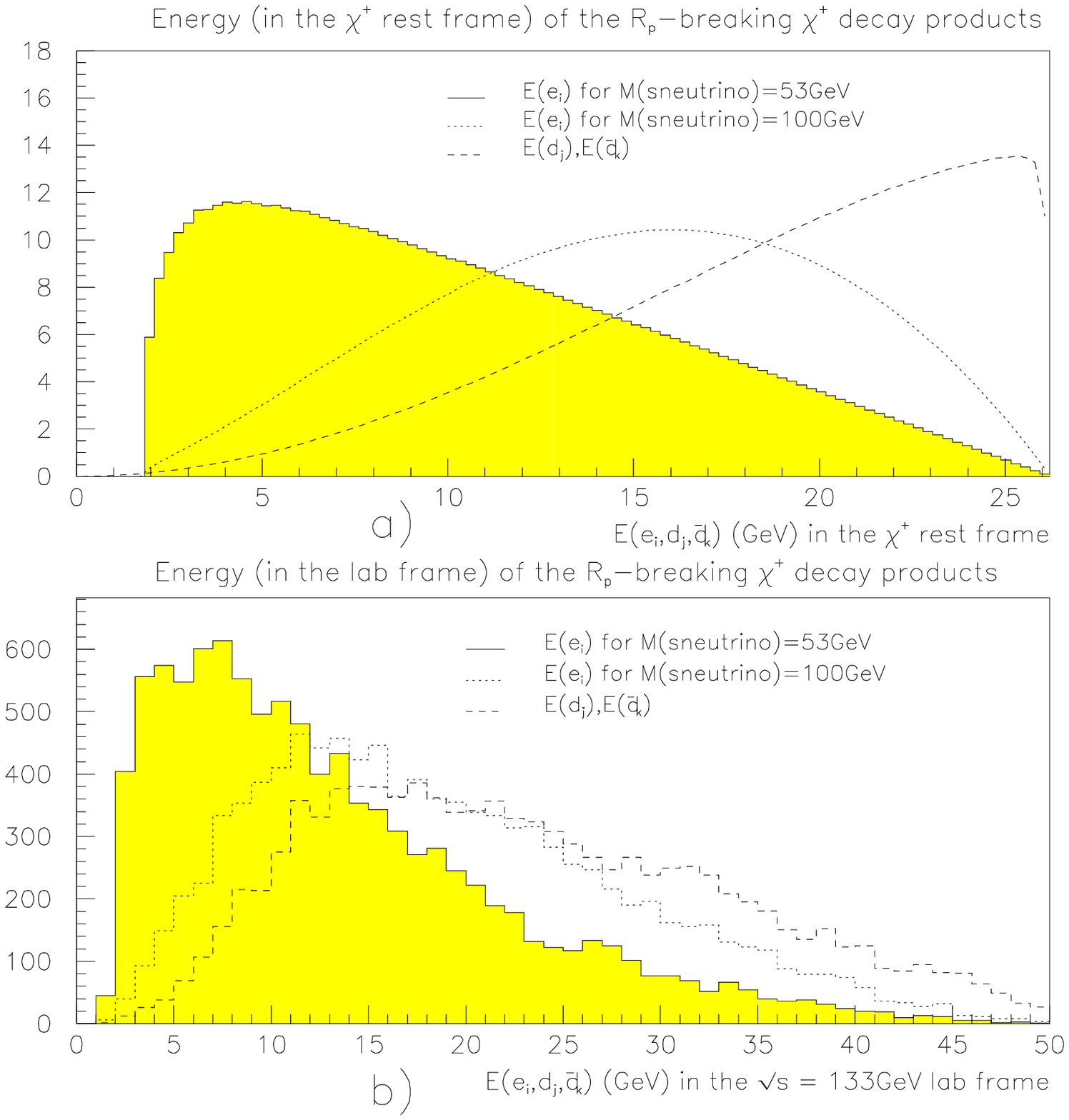}{9cm}{\footnotesize Energy distribution of the
$\chargino$ decay products. Here $\mu=54.5\gev$, $M_2=500\gev$, and $M_{{\tilde
\chi}+}=52.5\gev$. Plot a) shows the Energy in the $\chargino$ rest frame, plot
b) shows the Energy in the LEP lab frame.} {edis}

\section{Chargino Production and Signals}

We now turn to the $\rpv$ chargino signals at LEP. 
First we briefly mention the more conventional $\rpv$ signal from chargino
production, before we focus on the most
interesting signal, the direct $\rpv$ chargino decays which could explain the
recently observed excess in 4-jets by the ALEPH collaboration \cite{4jets} with
a combined invariant mass of $\sum M \approx 105\gev$.

\subsection{Chargino Cascade Decays and Neutralino LSP Models}

For a $\rpv$ Yukawa coupling $\lam'_{ijk}\lsim0.01$,  ${\Gamma_{MSSM}} \gg {
\Gamma_\rpvm}$ for most of the MSSM parameter space and the charginos will
dominantely decay (\rp-conserving) to the lighter neutralino via 
(\ref{eq:mssmcascades1})-(\ref{eq:mssmcascades2}). If the neutralino is the 
LSP it will decay \rp-violating to 
\barr
\label{eq:neutdec1}
\neutralino &\ra& e^+_i + u_j + {\bar d}_k \hspace {1cm} + h.c.\\
\label{eq:neutdec2}
\neutralino &\ra& {\nu}_i + d_j + {\bar d}_k \hspace {1.2cm} + h.c.
\earr
in the case of one dominant $L_i Q_j {\bar D}_k$ coupling. The neutralino decays
are discussed in detail in \cite{peterherbi}. The overall $\rpv$-signal for 
chargino pair production at LEP is then 
\barr
e^+ + e^- \ra {\tilde\chi}^+ + {\tilde\chi}^- \ra 2 \times  \left\{ 
\begin{array}{l}
 l + \nu + \neutralino \\
 q + {\bar q}' +  \neutralino 
\end{array}
\right.
\earr
and the neutralinos decay subsequently via
(\ref{eq:neutdec1})-(\ref{eq:neutdec2}).

\subsection{Direct $\rpv$ Chargino Decays and 4-Jet Signals}
\label{sec:4jets}

As we have seen (Fig.\ref{model}) the chargino decays \rp-violating directly
into SM particles even when it is not the LSP for Yukawa couplings in the range
$\lam_{ijk}' = 0.3 - 0.03$. Because strict limits on $\lam_{ijk}'$ from
low-energy constraints or from experimental direct searches exist, only the
couplings with the weakest bounds are of interest to this model: $\lam_{3jk}$ 
(see also Table \ref{lambdabounds}). For these couplings the topology of the
signal changes considerably compared to the ``conventional'' $\rpv$-signals
discussed in the previous subsection. 

In order to illustrate this we focus on the coupling $\lam_{3jk}$. We show how
$\rpv$ chargino decays could explain the excess of 4-jets seen by ALEPH
\cite{4jets} at a combined invariant mass of $\sum M \approx 105\gev$. Under a
chargino pair-production hypothesis the charginos would have a mass of
$M_{\chargino}=52.5\gev$. The cross-section is $\sigma_{{\tilde \chi}^{+}
{\tilde \chi}^{-}} \approx 6.7$pb for $M_{\chargino}=52.5\gev$\footnote{The
cross-section is fairly independent of the gaugino parameters along
$M_{\chargino}=52.5\gev$.} at ${\sqrt s}=133\gev$, compatible with the observed
excess of 4-jet events \cite{4jets}. Any set of gaugino parameters along the
chargino contour of $M_{\chargino}=52.5\gev$ within the grey region in
Fig.\ref{model} would furthermore allow for direct $\rpv$ chargino decays into
$\chargino \ra \tau^+ d_j {\bar d}_k$. And as we have already seen in section
\ref{Decay.Distributions} the tau energy distribution can be very soft. The
interpretation of ALEPH's excess in 4-jet events could thus be\footnote{
Indices $j,k$ are generation indices of the down-type quarks.} 
\beq
e^+ + e^- \ra {\tilde\chi}^+ + {\tilde\chi}^- \ra d_j {\bar d}_k {\bar d}_j d_k
\tau^+ \tau^-
\label{eq:4jetinter}
\eeq
where the taus are soft and mostly decay semi-hadronically. The overall final
state is experimentally reconstructed as a 4-jet final state. If the 4-jet
signal seen by ALEPH persists with higher statistics and more data  one could
easily verify this $\rpv$-model by looking for signs of high momentum leptons
from the tau decays. Note that although the tau energy spectrum is soft, when
boosted to the LEP lab frame ($\sqrt s = 133\gev$) the tau high energy tail
extends to  $40\gev$ or more - Fig.\ref{edis}b. Depending on the ratio of
$(m_{\tilde\nu}/m_{\tilde e})$, one also expects  chargino decays to the
mode (\ref{eq:finalstates}.1) resulting in missing momentum final states. 

Furthermore for a given set of gaugino parameters one would also expect
neutralino pair production. It turns out that along $M_{\chargino}=52.5\gev$
the neutralino cross-section $\sigma_{e^+ + e^- \ra {{\tilde \chi}^0_1} +
{{\tilde \chi}^0_2}} \approx 7.2$pb (at ${\sqrt s}=133\gev$) is fairly constant
and dominates over $\sigma_{{{\tilde \chi}^0_1} {{\tilde \chi}^0_1}}$ and
$\sigma_{{{\tilde \chi}^0_2} {{\tilde \chi}^0_2}}$ . Hence an additional signal
\beq
e^+ e^- \ra {{\tilde \chi}^0_1} + {{\tilde\chi}^0_2} \ra {{\tilde \chi}^0_1} + 
\left\{ \begin{array}{l} 
 {{\tilde \chi}^0_1} + q + {\bar q}\\
{{\tilde \chi}^0_1} +  l + {\bar l}\\
{{\tilde \chi}^0_1} +  \nu + {\bar \nu}\\
{{\tilde \chi}^0_1} +  \gamma
\end{array}
\right.
\eeq
would be expected.

\section{Conclusions}
We have calculated direct $\rpv$ chargino decays and have found that there are
regions in the MSSM parameter space in which the chargino will decay to $\tau^+
+ d_j +{\bar d}_k$ even when it is not the LSP for values of $\lam_{3jk}'$
below present experimental bounds. We therefore interpret the recently observed
excess in 4-jet events by ALEPH as chargino pair production with subsequent
$\rpv$ decays to 4 quarks and 2 soft taus, which are experimentally
reconstructed as 4-jets.  Further analysis has shown that the chargino
cross-section and the decay distributions are compatible with the observed
4-jet signal. 

The other suggestions to explain the four jet excess by sfermion pair production
\cite{rpsol1,rpsol2} are experimentally distinguishable from our interpretation.
We consider a heavier sfermion spectrum. The cross section for sneutrino
pair production is a factor of three lower than the squark pair production
which in turn is a bit lower than the chargino production rate. Furthermore
we have suggested $\geq4$ jet final states which have been experimentally
tagged as 4 jet final states. These four jets should thus typically be broader
than a true 4 jet event. More data is eagerly awaited to confirm or reject
our interpretation.

\section{Acknowledgements}
Many thanks to Michael Schmitt, Matthew Williams, Peter Dornan, Grahame Blair
and John Thompson for a number of discussions which this analysis has greatly benefitted
from. Furthermore we would like to thank the ALEPH collaboration, who has
motivated our work by providing the data and by giving a warm welcome to our
ideas and encouring our work.

\section{Appendix}
We here collect some formulas related to the amplitudes squared of the
chargino decay rate (\ref{eq:matrix1})-(\ref{eq:matrix4}). The squares of
the propagators are given in terms of the momenta and the sfermion masses
\barr
R({\tilde e}_{iL}) &=&(\chi^+-\nu_i)^2-{\tilde m}^2_{eiL},\quad
R({\tilde d}_{jL}) =(\chi^+- u_j)^2-{\tilde m}^2_{djL},\\
R({\tilde \nu}_{iL}) &=&(\chi^+-e_i)^2-{\tilde m}^2_{\nu iL}, \quad
R({\tilde u}_{jL}) =(\chi^+-d_j)^2-{\tilde m}^2_{ujL},\\
R({\tilde d}_{kR}) &=&(\chi^+-u_k)^2-{\tilde m}^2_{dkR}.
\earr
The coupling constants are given by
\barr
\alpha_L&=&0, \quad \alpha_R=-iU_{l1} \\
\beta_L&=&\frac{im_{uj}V^*_{l2}}{\sqrt{2}M_W\sin\beta}, \quad \beta_R=\alpha_R\\
\gamma_L&=&iV^*_{l1}, \quad \gamma_R=-\frac{igm_{ei}U_{l2}}
{\sqrt{2}M_W\cos\beta}\\
\delta_L&=&\gamma_L, \quad \delta_R= -\frac{igm_{dj}U_{l2}}
{\sqrt{2}M_W\cos\beta}
\earr
We had already factored out the $SU(2)$ coupling $g$ in the matrix elements.
We follow here the notation of \cite{gunhab}, where one can also find the 
expressions for the matrices $U_{ij},\,V_{ij}$ which diagonalize the
chargino mass matrix.

For the operator $L_iL_j{\bar E}_k$, ($i\not=j$) the chargino can decay
into the final states
\barr 
{{\tilde\chi}^+}_l \ra\left\{ \begin{array}{l}
{\nu}_i+\nu_j+e^+_{kR}   \,\,\,\,\,\,\,\,   (\ref{eq:llefinalstates}.1)   \\ 
e_i^++e_j^++e_{kR}^-      \,\,\,\,\,\,  (\ref{eq:llefinalstates}.2) 
\end{array}
\right.
\label{eq:llefinalstates}
\earr
The corresponding matrix elements squared are given by
\barr
\label{eq:llematrix1}
|{\cal M}_1|^2 &=& 4g^2{\lam}^2 \left[
\frac{\alpha_R^2}{R^2({\tilde e}_{iL})} \dpr{\charginol}{\nu_i} 
                     \dpr{\nu_j}{{\bar e}_k} 
+ \frac{\beta_R^2}{R^2({\tilde e}_{jL})}
\dpr{{\charginol}}{\nu_j}\dpr{\nu_i}{{\bar e}_k}\right.\nonumber \\
&&\left.-{\cal R}e\left\{
 \frac{ \alpha_R\beta_R^* } { R({\tilde e}_{iL})R({\tilde e}_{jL}) }
{\cal G}(p,\nu_i,{\bar e}_k,\nu_j)
\right\} \right] \\
\label{eq:llematrix2}
|{\cal M}_2|^2 &=&  
4g^2{\lam}^2 \left[\frac{\dpr{e_j}{{\bar e}_k}}{R^2({\tilde \nu}_{iL})}
\left\{(\gamma_L^2+\gamma_R^2)\dpr{{\charginol}}{e_i} 
+2{\cal R}e(\gamma_L\gamma_R^* m_{ei}M_{{\charginol}}
)\right\} \nonumber \right.\\
&& + \frac{\dpr{e_i}{{\bar e}_k}}{R^2({\tilde \nu}_{jL})}
\left\{(\delta_L^2+\delta_R^2)\dpr{{\charginol}}{e_j} 
+2{\cal R}e(\delta_L\delta_R^* m_{ej}M_{{\charginol}})
\right\}\nonumber \\
&&\left.-\frac{1}{R({\tilde \nu}_{iL})R({\tilde \nu}_{jL})} {\cal R}e\left\{
\gamma_L \delta_L^*{\cal G}({\charginol},e_i,{\bar e}_k,e_j)
+\gamma_L \delta_R^* m_{ej}M_{{\charginol}}\dpr{e_i}{{\bar e}_k} \right.\right.
\nonumber \\
&& \left.\left.
+\gamma_R \delta_L^* m_{ei}M_{{\charginol}}\dpr{e_j}{{\bar e}_k}
+\gamma_R \delta_R^* m_{ei}m_{ej}\dpr{{\charginol}}{{\bar e}_k} \right\}\right].
\earr
$\alpha,\beta,\gamma,\delta$ are given as above except that in $\delta_R$
$m_{dj}$ is replaced by $m_{ej}$ and $\beta_L=0$ because of vanishing neutrino
mass. Again we have included all mass effects. These are now only relevant for
large $\tan\beta$.

\end{document}